\documentclass[english,aps,reprint]{revtex4-1}
\usepackage[T1]{fontenc}
\usepackage[latin9]{inputenc}
\usepackage{color}
\usepackage{babel}
\usepackage{textcomp}
\usepackage{mathrsfs}
\usepackage{amsmath}
\usepackage{amssymb}
\usepackage{graphicx}
\usepackage{wasysym}
\usepackage[unicode=true,
 bookmarks=false,
 breaklinks=false,pdfborder={0 0 1},backref=section,colorlinks=false]
 {hyperref}

\makeatletter

\providecommand{\tabularnewline}{\\}

\usepackage{babel}

\makeatother

\begin{document}
\title{Quantum Diffusive Magneto-transport in Massive Dirac Materials with
Chiral Symmetry Breaking}
\author{Bo Fu}
\affiliation{Department of Physics, The University of Hong Kong, Pokfulam Road,
Hong Kong, China}
\author{Huan-Wen Wang}
\affiliation{Department of Physics, The University of Hong Kong, Pokfulam Road,
Hong Kong, China}
\author{Shun-Qing Shen}
\email{sshen@hku.hk}

\affiliation{Department of Physics, The University of Hong Kong, Pokfulam Road,
Hong Kong, China}
\begin{abstract}
Massive Dirac fermions break the chiral symmetry explicitly and also
make the Berry curvature of the band structure non-Abelian. By utilizing
the Green's function technique, we develop a microscopic theory to
establish a set of quantum diffusive equations for massive Dirac materials
in the presence of electric and magnetic fields. It is found that
the longitudinal magnetoresistance is always negative and quadratic
in the magnetic field, and decays quickly with the mass. The theory
is applicable to the systems with non-Abelian Berry curvature and
resolves the puzzles of anomalous magnetotransport properties measured
in topological materials.
\end{abstract}
\maketitle

\paragraph*{Introduction.}

Symmetries and their corresponding conservation laws play an important
role in understanding the fundamental nature of matter. However, a
classical conservation law might turn out to be violated in its quantized
version,\textit{ i.e.}, the so-called quantum anomaly \citep{BardeenW-69pr,fujikawa2004path}.
A well-known example is that the massless relativistic Dirac fermions
or Weyl fermions in three spatial dimensions possess the chiral anomaly
\citep{adler1969axial,Bell69nca}. In 1983, Nielsen and Ninomiya \citep{Nielsen83pl}
proposed that the chiral anomaly of the Weyl fermions could be realized
in the Weyl semimetals based on the picture of the Landau levels of
the Weyl fermions in a finite magnetic field as shown in Fig. 1(left),
and a negative magnetoresistance is regarded as a substantial signature
of the effect. Since then, a lot of theoretical approaches have been
developed for the anomaly-induced magnetoresistivity for massless
Weyl fermions \citep{stephanov2012chiral,Son13prb,bacsar2014triangle,burkov2014chiral,burkov2015negative,Lu-17fp}.
Recent advances in topological materials demonstrate a series of topological
materials may host the chiral quasi-particles \citep{Murakami-07njp,Wan-11prb,WangZJ-13prb,Xu15Science,HuangSM15natcomm,lv2015experimental,lu2015experimental,armitage2018weyl},
which provide a practical route to detect the signatures of the purely
quantum mechanical effect. The longitudinal negative magnetoresistance
has been reported experimentally in a large class of topological materials
\citep{kim2013dirac,huang2015observation,XiongJ15science,li2015giant,Zhang16natcomm,arnold2016negative,li2016negative,Liang2018,Li16natphys}.
However, a puzzle arises as some topological materials with negative
magnetoresistance are actually not Weyl semimetals: for example, $\mathrm{Zr}\mathrm{Te}{}_{5}$
\citep{Li16natphys,ZhangY-17nc,xu2018temperature,Mutch2019} and $\mathrm{Cd_{2}\mathrm{As_{3}}}$
\citep{Jeon-14nm} actually have a tiny direct band gap, and $\mathrm{Bi}{}_{2}\mathrm{Se}{}_{3}$
is a typical topological insulator \citep{WangJ-12nr,wiedmann2016anisotropic,Breunig2017}.
It is known that massive Dirac fermions break the chiral symmetry
explicitly. Another direct consequence of the chiral symmetry breaking
is that the Berry curvature of the band structure is non-Abelian \citep{shindou2005noncommutative,chang2008berry,chuu2010semiclassical,chen2014kinetic}.
Thus it becomes an open issue whether the measured negative magnetoresistance
could be still attributed to the chiral anomaly in the case of massive
Dirac fermions. Some mechanisms have been proposed for topological
and trivial states without invoking chiral anomaly \citep{goswami2015axial,gao2017intrinsic,dai2017negative,andreev2018longitudinal,wang2018intrinsic,fu2019interference}.
However, it is desirable to develop a unified quantum magnetotransport
theory for the topological materials with either Abelian or non-Abelian
Berry curvature to clarify the puzzle.

\begin{figure}
\includegraphics[width=8cm]{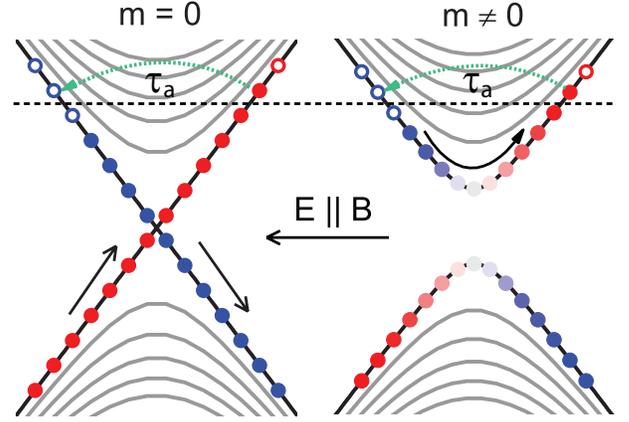}

\caption{Schematics for chiral anomaly related magnetoresistance mechanism
for massless (left panel) and massive (right panel) Dirac materials
in the parallel electromagnetic field. The occupied and unoccupied
states are shown as solid and open dots for the lowest Landau level,
respectively, and the color of the dots indicates the averaged chirality
$\langle\gamma^{5}\rangle$ as a function of the momentum. For massive
case of $m\protect\neq0$, the states are mixed near $k_{z}=0$. The
populations for two chiralities are different due to the presence
of parallel electric and magnetic field. The black solid line arrows
demonstrate the charge transfer driven by the electric field. The
exceeding right hand electrons (red solid dots) are scattered back
to left (blue open dots) as indicated by the dashed green line arrow,
and the relaxation time is charactered by $\tau_{a}$.}
\end{figure}
In this Letter, we develop a quantum diffusive theory for massive
Dirac materials in a finite uniform magnetic field by using the diagrammatic
perturbation theory. A set of the coupled diffusive equations is derived
for all the 16 relevant physical observables in terms of the Dirac
matrices. The quantum fluctuation for the axial charge and current
density still survives for massive Dirac fermions, and consequently,
the longitudinal magnetoconductivity is found to be positive and quadratic
in a magnetic field and decays quickly with the mass. A renormalized
continuity equation for the axial density and currents is obtained
in the presence of the electric and magnetic field. Our calculation
also demonstrates that the anomaly correction is rooted in the current
vertex renormalization from the axial charge density.

\paragraph*{Model Hamiltonian.}

We start with the Hamiltonian for massive Dirac fermions,
\begin{equation}
\mathscr{H}_{0}=\int d^{3}\mathbf{x}\bar{\Psi}(\mathbf{x})\left(v\hat{\mathbf{p}}\cdot\boldsymbol{\gamma}+\varepsilon_{F}\gamma^{0}+mv^{2}\right)\Psi(\mathbf{x}),\label{eq:H0}
\end{equation}
where $m$ is the Dirac mass, $v$ is the effective velocity, $\varepsilon_{F}$
is the chemical potential, and $\hat{\mathbf{p}}=-i\hbar\boldsymbol{\nabla}$
is the momentum operator. $\Psi(x)$ is the four-component Dirac spinor
with the time-space position four vector $x^{\mu}=(t,\mathbf{x})$,
that the Greek indices ($\mu,\nu,$ etc.) run over all the spacetime
indices ($0,1,2,3$). $\gamma^{\mu}$ are the Dirac gamma matrices
in Weyl representation $\gamma^{0}=\tau^{3}\otimes\sigma_{0}$ and
$\gamma^{i}=i\tau^{2}\otimes\sigma^{i}$ ($i=1,2,3$ ) with $\tau^{i}$
and $\sigma^{i}$ are the Pauli matrices, acting on the orbital and
spin degrees of freedom correspondingly. The chirality operator is
$\gamma^{5}=i\gamma^{0}\gamma^{1}\gamma^{2}\gamma^{3}=\tau^{1}\otimes\sigma^{0}$.
The mass term breaks the chiral symmetry, and also modifies the Nielsen-Ninomiya's
picture for chiral anomaly at a finite field. The two 0th Landau levels
are mixed together near the crossing point as shown in Fig. 1(right).
The charge tunneling process can be realized through the smoothly
connected region, which is analogous to the massless case that the
chiral charge pumping is through the infinite Dirac sea \citep{Nielsen83pl}.
The conservation law for the axial charge is modified to be \citep{zee2010quantum}

\begin{equation}
\partial_{\mu}\hat{J}^{a\mu}(x)=2mv^{2}\hat{\mathfrak{n}}_{P}+\frac{e^{3}}{2\pi^{2}\hbar^{2}}\mathbf{E}\cdot\mathbf{B}\label{eq:chiralanomaly}
\end{equation}
where $\hat{\mathfrak{n}}_{P}=\bar{\Psi}i\gamma^{5}\Psi$ is the pseudo-scalar
density, indicating that the axial charges are even not conserved
at the classical level in the presence of the Dirac mass. Furthermore,
the anomaly term arises as a consequence of the ultraviolet divergence
of the ``VVA'' triangle diagrams which cannot be cured by a finite
mass \citep{zee2010quantum}. In a uniform magnetic field, say along
the z-direction, the kinetic momentum operator $\hat{\mathbf{p}}$
is replaced by the canonical momentum operator in Eq. (\ref{eq:H0}),
$\boldsymbol{\pi}=-i\hbar(\boldsymbol{\nabla}-ie\mathbf{A})$ with
the gauge field chosen as $\mathbf{A}=(-Bx_{2},0,0)$. In the case
the model is solvable, and the energy dispersion becomes discrete
to form the Landau levels \citep{wang2018intrinsic}. The Green's
functions for the free Dirac fermions at a magnetic field can be obtained
analytically {[}see Sec. S1 in \citep{Note-on-SM}{]}.

\begin{table}
\begin{tabular}{|c|c|c|c|c|c|}
\hline
bilinear ($\gamma^{\mathtt{A}}$)  & physical quantity  & $\mathcal{T}$  & $\mathcal{I}$  & $\mathcal{C}$  & disorder \tabularnewline
\hline
\hline
$\bar{\Psi}\gamma^{0}\Psi$  & total charge $(J^{0})$  & $\checked$  & $\checked$  & $\checked$  & $\Delta$\tabularnewline
\hline
$\bar{\Psi}\gamma^{0}\gamma^{5}\Psi$  & axial charge $(J^{a0})$  & $\checked$  & $\times$  & $\checked$  & $\Delta_{a}$\tabularnewline
\hline
$\bar{\Psi}\Psi$  & scalar mass $(\mathfrak{n}_{\beta})$  & $\checked$  & $\checked$  & $\times$  & $\Delta_{m}$\tabularnewline
\hline
$\bar{\Psi}i\gamma^{5}\Psi$  & pseudo-scalar density $(\mathfrak{n}_{P})$  & $\times$  & $\times$  & $\times$  & $\Delta_{P}$\tabularnewline
\hline
$\bar{\Psi}\gamma^{i}\Psi$  & current $(J^{i})$  & $\times$  & $\times$  & $\checked$  & $\Delta_{c}$\tabularnewline
\hline
$\bar{\Psi}\gamma^{5}\gamma^{i}\Psi$  & axial current $(J^{ai})$  & $\times$  & $\checked$  & $\checked$  & $\Delta_{ac}$\tabularnewline
\hline
$\bar{\Psi}i\gamma^{0}\gamma^{i}\Psi$  & electric polarization $(\mathfrak{p}_{i})$  & $\checked$  & $\times$  & $\times$  & $\Delta_{p}$\tabularnewline
\hline
$\bar{\Psi}\gamma^{5}\gamma^{0}\gamma^{i}\Psi$  & magnetization $(\mathfrak{m}_{i})$  & $\times$  & $\checked$  & $\times$  & $\Delta_{M}$\tabularnewline
\hline
\end{tabular}

\caption{Various types of physical quantities and disorder represented by fermionic
bilinears ($i=1,2,3$), their symmetries under time-reversal ($\mathcal{T}$),
parity ($\mathcal{I}$), and continuous chiral rotation ($\mathcal{C}$).
The time-reversal symmetry $\mathcal{T}$ is generated by an anti-unitary
operator $\gamma^{1}\gamma^{3}\mathcal{K}$, where $\mathcal{K}$
is complex conjugation, such that $\mathcal{T}^{2}=-1$ . The parity
operator is generated $\mathcal{I}=\gamma^{0}$. The continuous chiral
symmetry ($\mathcal{C}=e^{i\theta\gamma^{5}}$) is generated by $\gamma^{5}$.
Here $\checked$ and $\times$ signifies even and odd under a symmetry
operation, respectively. And, we use the Latin capital letters $\mathtt{A},\mathtt{B},...$
for indices when the index runs through the entire hypercomplex system
from $1$ to $16$.}
\end{table}
By using the five Dirac gamma matrices $\gamma^{\mu}$ ($\mu=0,1,2,3$
and $5$) and their descendants, we can define 16 physical quantities
as shown in Table I. The enlarged (pseudospin $\otimes$ spin) gamma
matrices will allow us to obtain a microscopic theory of diffusive
transport for all the possible coupled physical observables in the
presence of an external field. The Dirac structure of the Hamiltonian
also allows various types of disorder. For simplicity, we only concentrate
on impurities with time reversal and parity invariance
\begin{equation}
\mathscr{H}_{dis}=\int d^{3}\mathbf{x}\left[V(\mathbf{x})\bar{\Psi}(\mathbf{x})\gamma^{0}\Psi(\mathbf{x})+V_{m}(\mathbf{x})\bar{\Psi}(\mathbf{x})\Psi(\mathbf{x})\right],
\end{equation}
which corresponds to the random chemical potential and mass respectively.
All the disorders are quenched, random variables behaving as white
noises, $\langle\langle V(\mathbf{x})V(\mathbf{x}^{\prime})\rangle\rangle=\Delta\delta^{3}(\mathbf{x}-\mathbf{x}^{\prime})$
and $\langle\langle V_{m}(\mathbf{x})V_{m}(\mathbf{x}^{\prime})\rangle\rangle=\Delta_{m}\delta^{3}(\mathbf{x}-\mathbf{x}^{\prime})$,
and further assume the random chemical potential dominates the elastic
scattering processes $\Delta\gg\Delta_{m}$. The self-energy is calculated
in the Born approximation and we neglect its magnetic field dependence.
The real part of the self-energy gives a shift of the zero of energies
and will not be considered either. The imaginary part of self-energy
can be evaluated as $\mathrm{Im}\Sigma^{R}=-\frac{\pi\rho}{2}(\Delta+\Delta_{m})(\gamma^{0}+\eta\boldsymbol{1})$,
where $\rho=k_{F}^{2}/(2\pi^{2}\hbar v_{F})$ is the density of states
with Fermi wavevector $k_{F}=\sqrt{\varepsilon_{F}^{2}-m^{2}v^{4}}/\hbar v$
and Fermi velocity $v_{F}=(\hbar v^{2}k_{F})/\varepsilon_{F}$. The
orbital polarization is defined by $\eta\equiv\langle\gamma^{0}\rangle=mv^{2}/\varepsilon_{F}$.
Thus the total relaxation time (or quasiparticle lifetime) is given
by $\tau=\hbar/[\pi\rho(\Delta+\Delta_{m})(1+\eta^{2})]$ and the
random mass induced relaxation time is $\tau_{m}=\hbar/[\pi\rho\Delta_{m}(1+\eta^{2})]$.

\paragraph*{Quantum diffusive equations in the real space.}

With the help of Table I, we can introduce the 16-dimensional vectors
$\hat{\mathcal{S}}_{\mathtt{A}}(x)=d_{\mathtt{A}}\bar{\Psi}(x)\gamma^{\mathtt{A}}\Psi(x)$
for all the possible physical observables in terms of the Dirac matrices,
where $d_{\mathtt{A}}=e$ for the charge density operators (the first
four quantities in Table 1) and $d_{\mathtt{A}}=ev$ for the current
operators. To investigate the response to the external potential $\mathcal{A}_{\mathtt{A}}(x)$
, we consider the generic external perturbation$\mathscr{H}_{1}(t)=\sum_{\mathtt{A}}\int d^{3}\mathbf{x}\hat{\mathcal{S}}_{\mathtt{A}}(x)\mathcal{A}_{\mathtt{A}}(x)$
\citep{mahan2013many} with $\mathcal{A}_{\mathtt{A}}$ are also 16-dimensional
vectors. The observables can be evaluated within the framework of
the linear response theory \citep{Kubo-57jspj}, $\mathcal{S}_{\mathtt{A}}\approx\mathcal{S}_{\mathtt{A}}^{(0)}+\mathcal{S}_{\mathtt{A}}^{(1)}+\mathcal{O}(\mathcal{A}_{\mathtt{A}}^{2})$.
$\mathcal{S}^{(0)}=-d_{\mathtt{A}}\mathrm{Tr}[\gamma^{\mathtt{A}}G(x,x)]$
is the zeroth order term in $\mathscr{H}_{1}$ and $G(x,x^{\prime})$
is the fermion propagator for $\mathscr{H}_{0}$. $S_{\mathtt{A}}^{(1)}(x)=\int d^{4}x^{\prime}\chi_{\mathtt{A}\mathtt{B}}^{R}(x,x^{\prime})\mathcal{A}_{\mathtt{B}}(x^{\prime})$
is the first order response to $\mathscr{H}_{1}$, with $\chi_{\mathtt{A}\mathtt{B}}^{R}(x,x^{\prime})$
is the retarded response function which can be evaluated by analytical
continuation of the imaginary time expression. In order to establish
a set of the diffusion equations to describe the coupled dynamics
of all the physical quantities, we need to consider the vertex renormalization
due to the multi-scattering which appears perturbatively as a series
of impurity line ladder diagrams, in which only combinations of retarded
and advanced Green's functions, having poles on opposite sides of
the real axis, will contribute. As a consequence, the bare vertex
$\gamma^{\mathtt{B}}$ in the linear's response theory should be replaced
by the dressed vertex $\varGamma^{\mathtt{B}}(\mathbf{x},\mathbf{x}^{\prime};\omega)$,
which satisfies the Bethe-Salpter equation. By expanding the renormalized
vertex $\varGamma^{B}$ in terms of the Dirac matrix $\gamma^{C}$,
we yield a $16\times16$ matrix $\widetilde{\varGamma}$ with its
elements defined as $\widetilde{\varGamma}^{\mathtt{B}\mathtt{C}}=\frac{1}{4}\mathrm{Tr}\Big[\varGamma^{B}\gamma^{\mathtt{C}}\Big]$.
In the diffusive regime or hydrodynamic regime, the spatial variations
of $\widetilde{\varGamma}(\mathbf{x},\mathbf{x}^{\prime};\omega)$
are small on the scale of the mean free path $\ell_{e}=v_{F}\tau$,
then we expand $\widetilde{\varGamma}(\mathbf{x}_{1},\mathbf{x}^{\prime};\omega)$
about $\mathbf{x}_{1}=\mathbf{x}$: $\widetilde{\varGamma}(\mathbf{x}_{1},\mathbf{x}^{\prime};\omega)\approx\widetilde{\varGamma}(\mathbf{x},\mathbf{x}^{\prime};\omega)+(\mathbf{x}_{1}-\mathbf{x})\cdot\nabla_{\mathbf{x}}\widetilde{\varGamma}(\mathbf{x},\mathbf{x}^{\prime};\omega)+\frac{1}{6}(\mathbf{x}_{1}-\mathbf{x})^{2}\nabla_{\mathbf{x}}^{2}\widetilde{\varGamma}(\mathbf{x},\mathbf{x}^{\prime};\omega)$
and substitute it into the Bethe-Salpter equation. Finally, we obtain
the diffusive equations for $\widetilde{\varGamma}(\mathbf{x},\mathbf{x}^{\prime};\omega)$,
$\mathscr{D}_{\mathbf{x}}^{-1}\widetilde{\varGamma}^{T}(\mathbf{x},\mathbf{x}^{\prime};\omega)=\mathcal{W}^{-1}\delta(\mathbf{x}-\mathbf{x}^{\prime})$
with $\mathscr{D}_{\mathbf{x}}^{-1}$ is the diffusion operator and
$\mathcal{W}$ is an impurity related diagonal matrix \citep{Note-on-SM}.
The linear response $\mathcal{S}^{(1)}(\mathbf{x},\omega)$ can be
expressed in terms of $\widetilde{\varGamma}(\mathbf{x},\mathbf{x}^{\prime};\omega)$.
By using the explicit form of the dressed retarded and advanced Green's
functions in the Landau level representation $G^{R/A}$\citep{miransky2015quantum},
we can calculate the full 16 \texttimes{} 16 matrix $\mathscr{D}^{-1}$
{[}see Sec S2 and S3 in\citep{Note-on-SM}{]}. In this way, we reach
at a set of quantum diffusive equations,

\begin{equation}
\mathscr{D}_{\mathbf{x}}^{-1}\mathcal{S}^{(1)}(\mathbf{x},\omega)=-\frac{2}{\pi}\left[\mathcal{W}^{-1}-\mathscr{D}_{\mathbf{x}}^{-1}\right]\mathcal{W}^{-1}i\omega\mathcal{A}(\mathbf{x},\omega).\label{eq:totalequation}
\end{equation}

\paragraph*{The longitudinal magnetoresistance.}

A general solution of the quantum diffusive equations is quite complicated.
As an application to explore the longitudinal magnetoresistance, we
focus on the linear response for the electric field also along the
z-direction. The perturbation part of the Hamiltonian is $\mathscr{H}_{1}(t)=\int d^{3}\mathbf{x}A_{3}(t)\bar{\Psi}(x)\gamma^{3}\Psi(x)$.
In this case, the diffusion operator $\mathscr{D}^{-1}$ can be reduced
into a block diagonal form: among all the 16 physical quantities,
only 4 observables we are interested in are coupled together in the
quantum diffusive regime.\textcolor{red}{{} }Thus we extract the
following coupled $4\times4$ sub-matrix of $\mathscr{D}^{-1}$ which
is spanned by $(J^{0},J^{a0},\mathfrak{n}_{\beta},J^{3})$:
\begin{widetext}
\begin{align}
\frac{\hbar}{\pi\rho\tau^{2}}\mathscr{D}_{4\times4}^{-1} & =\left(\begin{array}{cccc}
i\omega+\frac{\varLambda_{0}}{\tau}-\mathcal{D}\partial_{z}^{2} & \varUpsilon v\partial_{z} & \eta(i\omega-\frac{1}{\tau}-\mathcal{D}\partial_{z}^{2}) & \frac{1}{3}(1-\eta^{2})v\partial_{z}\\
\varUpsilon v\partial_{z} & (1-\eta^{2})(i\omega+\frac{\varLambda_{a0}}{\tau}-\mathcal{D}\partial_{z}^{2}) & \frac{\eta}{2}\varUpsilon v\partial_{z} & -\frac{1}{\tau}\varUpsilon\\
\eta(i\omega-\frac{1}{\tau}-\mathcal{D}\partial_{z}^{2}) & \frac{\eta}{2}\varUpsilon v\partial_{z} & \eta^{2}(i\omega+\frac{\varLambda_{\beta}}{\tau}-\mathcal{D}\partial_{z}^{2}) & \frac{1}{3}\eta(1-\eta^{2})v\partial_{z}\\
\frac{1}{3}(1-\eta^{2})v\partial_{z} & -\frac{1}{\tau}\varUpsilon & \frac{1}{3}\eta(1-\eta^{2})v\partial_{z} & \frac{1}{3}(1-\eta^{2})(i\omega+\frac{\varLambda_{3}}{\tau}-\mathcal{D}\partial_{z}^{2})
\end{array}\right).\label{eq:66De}
\end{align}
The coefficient $\varUpsilon\equiv\frac{1}{2}(\frac{\hbar v}{\varepsilon_{F}\ell_{B}})^{2}$
with the magnetic length $\ell_{B}=\sqrt{\hbar/eB}$ and the dimensionless
diffusion channel relaxation rates are: $\varLambda_{0}=\eta^{2}$,
$\varLambda_{\beta}=\frac{1}{\eta^{2}}$, $\varLambda_{3}=2\frac{(\Delta+2\Delta_{m})+\eta^{2}(2\Delta+\Delta_{m})}{(\Delta-\Delta_{m})(1-\eta^{2})}$
and the axial relaxation rate is $\varLambda_{a0}=2\frac{\Delta_{m}+\Delta\eta^{2}}{(\Delta-\Delta_{m})(1-\eta^{2})}$
from the chiral symmetry breaking. By substituting (\ref{eq:66De})
into Eq. (\ref{eq:totalequation}) and using the explicit form of
$\mathcal{W}=\Delta\boldsymbol{1}_{4}+\Delta_{m}\mathrm{diag}(1,-1,1,-1)$
for this $4\times4$ case, we arrive at the coupled charge-current
dynamics equations in the presence of electromagnetic field. By transforming
into frequency-momentum space, the coupled charge-current equations
can be solved in the diffusive regime ($\omega\tau,\mathcal{D}\tau q_{z}^{2}\ll1$):
\end{widetext}

\begin{align}
\delta J^{a0} & =\frac{\varepsilon_{F}}{v\hbar k_{F}}\frac{i\omega}{i\omega+\mathcal{D}^{*}q_{z}^{2}}\frac{e^{3}}{\hbar^{2}}\frac{EB\tau_{a}}{2\pi^{2}};\label{eq:chiralimbalance}\\
\delta J^{3} & =\frac{i\omega\sigma_{D}E}{i\omega+\mathcal{D}^{*}q_{z}^{2}}\left(1+\frac{3}{4}\frac{\tau_{a}}{\tau^{*}}\frac{1}{k_{F}^{4}\ell_{B}^{4}}\frac{i\omega}{i\omega+q_{z}^{2}\mathcal{D}^{*}}\right);\label{eq:variationofcurrent}\\
\delta J^{0} & =\frac{iq_{z}\sigma_{D}E}{i\omega+\mathcal{D}^{*}q_{z}^{2}}\left(1+\frac{3}{4}\frac{\tau_{a}}{\tau^{*}}\frac{1}{k_{F}^{4}\ell_{B}^{4}}\frac{i\omega}{q_{z}^{2}\mathcal{D}^{*}+i\omega}\right);\label{eq:variationofdensity}\\
\delta\mathfrak{n}_{\beta} & =\eta\delta J^{0},
\end{align}
where $\mathcal{D}^{*}=\frac{3}{2}\frac{(1+\eta^{2})(\Delta+\Delta_{m})}{(\Delta+2\Delta_{m})+\eta^{2}(2\Delta+\Delta_{m})}\mathcal{D}$
is the renormalized diffusion coefficient with the classical diffusion
constant $\mathcal{D}=v_{F}^{2}\tau/3$, and $\sigma_{D}=2e^{2}\rho\mathcal{D}^{*}$
is the Drude conductivity. We have introduced the ratio $\frac{\tau_{a}}{\tau^{*}}=\frac{1}{\varLambda_{a0}}\frac{\mathcal{D}^{*}}{\mathcal{D}}$,
where the axial relaxation time $\tau_{a}$ describes the attenuation
time of the axial charge when it propagates in the disordered medium
and $\tau^{*}=\tau(1+\eta^{2})$.

The solution in Eq. (\ref{eq:variationofcurrent}) gives the dynamic
longitudinal conductivity $\sigma_{zz}(\omega,\mathbf{q},B)=\delta J^{3}(\omega,\mathbf{q},B)/E(\omega,\mathbf{q})$
in a finite magnetic field. In the ``slow limit'', $\lim_{q\to0}\lim_{\omega\to0}(\omega/q_{z}^{2}\mathcal{D}^{*})\to0$
\citep{zlatic2014modern}, such that the perturbing potential is nearly
constant on the timescale $1/q_{z}^{2}\mathcal{D}^{*}$. Consequently,
in the thermodynamic equilibrium no current will be generated: $\lim_{q\to0}\lim_{\omega\to0}\sigma_{zz}(\omega,\mathbf{q},B)=0,$
which is also a requirement of the gauge invariance that a purely
longitudinal and static vector potential cannot induce any physical
current. In the ``rapid limit'', $\lim_{\omega\to0}\lim_{q\to0}(q_{z}^{2}\mathcal{D}^{*}/\omega)\to0$
\citep{zlatic2014modern}. In this case, we obtain the remarkable
result, $\lim_{\omega\to0}\lim_{q\to0}\sigma_{zz}(\omega,\mathbf{q},B)=\sigma_{D}+\sigma_{CA}(\eta,B)$,
with the anomaly-induced magnetoconductivity as
\begin{align}
\sigma_{CA}(\eta,B) & =\frac{3}{4}\frac{\tau_{a}}{\tau^{*}}\frac{1}{k_{F}^{4}\ell_{B}^{4}}\sigma_{D}=\frac{3}{16}\frac{\tau_{a}}{\tau^{*}}\left(\frac{B}{B_{F}}\right)^{2}\sigma_{D}.\label{eq:anomalyconductivity}
\end{align}
The magnetoconductivity is positive and quadratic in $B$ and $B_{F}=\frac{\hbar}{2e}k_{F}^{2}$.
In the massless case of $\eta=0$, only the impurities which break
the chiral symmetry (the impurity matrix anticommutes with the chiral
symmetry operator) can cause the scattering between different chiralities
(nodes) control the axial relaxation time, and we may reproduce the
previous results from the semiclassical theory $\lim_{\eta\to0}\sigma_{CA}(\eta,B)=\frac{3}{4\pi^{2}2\rho}(\frac{eB}{\hbar})^{2}\frac{\tau_{a}}{\tau}\frac{\hbar v}{\varepsilon_{F}^{2}}\sigma_{D}$
\citep{Son13prb}. In a massive case of $\eta\ne0$, the chiral symmetry
is broken explicitly due to the Dirac mass $m$, and the eigenstates
near the Fermi level mix the chiralities. The disorder with the chiral
symmetry (e.g. the chemical potential randomness) can cause the backscattering
between opposite helicity, giving an axial relaxation time proportional
to the inverse of $\eta^{2}$. As shown in Eq. (\ref{eq:anomalyconductivity}),
when the carrier density of the system is fixed, the relative magnetoconductivity
$\sigma_{CA}(\eta,B)/\sigma_{D}$ is only determined by two parameters,
the relative impurity strength $\tau/\tau_{m}=\Delta_{m}/(\Delta+\Delta_{m})$
and the orbital polarization $\eta$. As shown in Fig. 2(a), when
the relative impurity strength $\tau/\tau_{m}$ is fixed, the relative
anomaly-induced magnetoconductivity $\sigma_{CA}(\eta)/\sigma_{CA}(\eta=0)$
is suppressed as the parameter $\eta$ grows. When $\tau/\tau_{m}$
goes to zero, the mass becomes dominant in chiral symmetry breaking,
and the anomaly-induced magnetoconductivity quenches to zero more
quickly as $\eta$ grows. We also plot the absolute value of $\sigma_{CA}$
by using the realistic parameters according to Ref. \citep{Li16natphys}.
As shown in Fig. 2 (b), $\sigma_{CA}$ is strongly suppressed as $\tau/\tau_{m}$
and $\eta$ grows. Experimental observation of the anomaly-induced
magnetoconductivity requires a long axial current relaxation time
$\tau_{a}$, which stems from the near conservation of chiral charge,
and a lower carrier density. A finite mass cannot forbid such an effect,
but only suppress its contribution. From Eq. (\ref{eq:chiralimbalance}),
in the ``rapid limit'', the chirality imbalance $\delta J^{a0}=\frac{e^{3}}{\hbar^{2}}\frac{\varepsilon_{F}}{v\hbar k_{F}}\frac{EB\tau_{a}}{2\pi^{2}}$
is also self-consistently obtained, further confirming our calculations.
The chiral anomaly-induced magnetoconductivity is rooted in the current
vertex renormalization from the axial charge density in the presence
of parallel electromagnetic field {[}see Sec. S5 in \citep{Note-on-SM}{]}.
Therefore, we could not obtain such an anomaly correction only in
the Drude approximation by considering the bubble diagram.

\begin{figure}
\includegraphics[width=9cm]{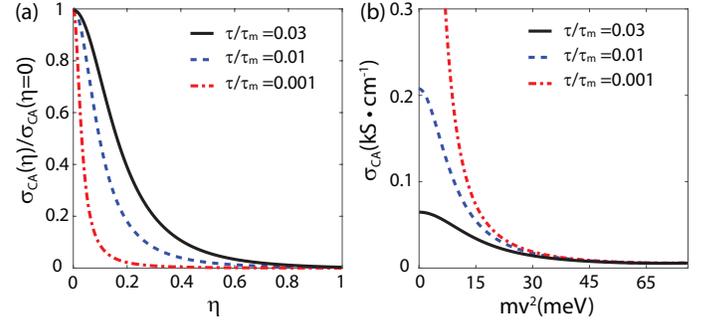}

\caption{(a) The universal behavior of the relative anomaly magnetoconductivity
correction for different $\tau/\tau_{m}$ as a function of the parameter
$\eta=mv^{2}/\varepsilon_{F}$. (b) The anomaly-related positive magneto
conductivities at $B=1T$ are plotted as a function of Dirac mass
$mv^{2}$ for several different $\tau/\tau_{m}$with the chemical
potential $\varepsilon_{F}=90\mathrm{meV}$ and Ohm resistance at
zero field $\mathcal{R}=\sigma^{-1}=1.2\text{\ensuremath{\mathrm{m\Omega\cdot cm}}}$.
The other parameter is chosen as $\hbar v=6\times10^{-5}\mathrm{meV\cdot cm}$.}
\end{figure}

\paragraph*{Anomaly induced magnetic field correction to the diffusive motion
of the electrons.}

The conservation of total charge also makes an anomaly-induced correction
to the dynamical polarization function or the density-density response
function $\chi_{00}(\omega,\mathbf{q},B)$. The gauge invariance poses
some constraints on the elements of the response function: $\chi_{00}=-\frac{q_{z}}{\omega}\chi_{03}$
with $\chi_{03}=\frac{\delta J^{0}}{A_{3}}$ , we yield the following
compact form for the polarization function from Eq. (\ref{eq:variationofdensity}),
\begin{equation}
\chi_{00}(\omega,\mathbf{q},B)=2\rho\frac{q_{z}^{2}\widetilde{D}(B)}{q_{z}^{2}\widetilde{D}(B)+i\omega},\label{eq:polarization}
\end{equation}
where $\widetilde{D}(B)=\mathcal{D}^{*}(1+\frac{3}{4}\frac{\tau_{a}}{\tau^{*}}\frac{1}{k_{F}^{4}\ell_{B}^{4}})$
is the field-dependent diffusion constant. This factor ($1/[q_{z}^{2}\widetilde{D}(B)+i\omega]$)
is known as the "diffusion pole", which emerges from the repeated
elastic scattering (the ladder diagram), and also reflects the conservation
of total charge. Many many-body effects are directly associated with
this diffusion pole. For example, when the electron-electron interaction
cannot be neglected, each electron will be influenced by the electronic
density fluctuation from other electrons described by $\chi_{00}$.
As a consequence, the spectral and transport properties are modified
by the interaction effect. One way to detect the effect is to measure
the tunnel conductance, which directly reflects the variation of the
density of states $\delta\rho$ due to the Coulomb interaction. The
reduction of the tunnel conductance is given by $\delta G_{t}(V)/G_{t}=\delta\rho(V)/\rho\propto(\sqrt{|eV|/\widetilde{D}(B)}-C)/\widetilde{D}(B)$
where $V$ is the voltage difference between two leads and $C$ is
a constant independent of the bias \citep{altshuler1985electron}.
Since the change is maximal around the Fermi energy $\varepsilon_{F}$,
the tunneling spectrum will display a downward cusp at the Fermi level,
\textit{i.e.}, the so-called zero-bias anomaly. Due to the magnetic
field dependence of the diffusion constant, we can expect the zero
bias downward cusp should be weakened under the magnetic field. Furthermore,
this interaction correction in conductivity shows a strong dependence
on the configuration of the electric and magnetic field in sharp contrast
with the contribution from weak localization, providing a fruitful
way to distinguish the two effects.

\paragraph*{Discussion and conclusion}

The psuedo-scalar density $\hat{\mathfrak{n}}_{P}$ modifies directly
the continuity equation for the axial charge and current density in
Eq. (\ref{eq:chiralanomaly}). In the presence of the electric and
magnetic field, it is found that the expectation value of the pseudo-scalar
density has the form $\langle\hat{\mathfrak{n}}_{P}\rangle=\frac{1}{2mv^{2}}(\frac{\varepsilon_{F}}{\hbar vk_{F}}-1)\frac{e^{3}}{2\pi^{2}\hbar^{2}}\mathbf{E}\cdot\mathbf{B}$,
which vanishes when $m=0$ \citep{fang2017pseudoscalar,lin2018mass,Note-on-SM}.
Thus the anomaly equation is reduced to
\begin{equation}
\partial_{\mu}J^{a\mu}=\frac{\varepsilon_{F}}{\hbar vk_{F}}\frac{e^{3}}{2\pi^{2}\hbar^{2}}\mathbf{E}\cdot\mathbf{B}.\label{eq:massivechiralanomaly}
\end{equation}
As the chemical potential $\varepsilon_{F}=\sqrt{\left(v\hbar k_{F}\right)^{2}+m^{2}v^{4}}$,
the prefactor $\frac{\varepsilon_{F}}{\hbar vk_{F}}=\sqrt{1+\left(\frac{mv}{\hbar k_{F}}\right)^{2}}$is
always larger than 1 for a finite mass. Assume the electric field
is caused by a spatially varying chemical potential $\varepsilon_{F},$
$\mathfrak{\mathbf{E=-\nabla}}\frac{\varepsilon_{F}}{e}$. Integrating
the anomaly equation leads to the dissipationless axial current $\mathbf{J}^{a}=-e\mathbf{B}v\hbar k_{F}\frac{e^{2}}{2\pi^{2}\hbar^{2}}$
for massive Dirac fermions\citep{gorbar2013radiative}, \textit{i.e.},
the so-called the chiral separation effect. The axial current will
induce a chiral charge separation, \textit{\textcolor{black}{i.e.,}}
a nonzero chiral chemical potential $\mu_{5}$. In a homogeneous case,
the chiral charge transfer rate is $\partial_{t}J^{a0}=\frac{\varepsilon_{F}}{\hbar vk_{F}}\frac{e^{3}}{2\pi^{2}\hbar^{2}}\mathbf{E}\cdot\mathbf{B}$
and the chiral chemical separation becomes nonzero $\mu_{5}$. The
energy cost for the energy transferring $\partial_{t}J^{a0}\mu_{5}$
should be supplied by the Joule heating, $\mathbf{J}\cdot\mathbf{E}$
\citep{SQS}. In this way it follows that $\mathbf{J}=\frac{\varepsilon_{F}}{\hbar vk_{F}}\frac{e^{2}}{2\pi^{2}\hbar^{2}}\mu_{5}\mathbf{B}$,
\textit{\textcolor{black}{i.e.}}, the chiral magnetic effect for massive
Dirac fermions. In the relaxation time approximation, we can obtain
the anomaly-induced magnetoconductivity in Eq. (\ref{eq:anomalyconductivity}).

In short, we have established a set of the quantum diffusive equations
for massive Dirac fermions. As an application, we calculated the longitudinal
magnetoresistance and the dynamical polarization functions. The longitudinal
magnetoresistance is always negative for both massive Dirac fermions
and Weyl fermion and decays quickly with the mass or energy gap. It
resolves the puzzle of the longitudinal negative magnetoresistance
observed in several topological Dirac semimetals and even in topological
insulators. Finally, it is worth stressing that this quantum diffusive
theory is applicable to the systems with either the Abelian or non-Abelian
Berry curvature.
\begin{acknowledgments}
This work was supported by the Research Grants Council, University
Grants Committee, Hong Kong under Grant No. 17301116.
\end{acknowledgments}

\bibliographystyle{apsrev4-1}

\end{document}